\documentclass[12pt]{article}
\usepackage[dvipdfmx]{graphicx}
\usepackage{subfigure}
\usepackage{amsmath}
\usepackage{mathptmx}     
\usepackage{latexsym}
\usepackage{amssymb}

\begin{document}
\title{Information cascade  on    networks}

\author{Masato Hisakado\footnote{[1]
masato.hisakado@fsa.go.jp}
 \space{} and   Shintaro Mori\footnote{[2] mori@sci.kitasato-u.ac.jp} }

\maketitle

\maketitle

*Financial Services Agency, Kasumigaseki 1-6-5, Chiyoda-ku, Tokyo 100-8967, Japan

\vspace*{1cm}

\dag Department of Physics, School of Science,
Kitasato University, Kitasato 1-15-1, Sagamihara, Kanagawa 252-0373, Japan

\vspace*{1cm}

 \vspace*{1cm}

\begin{abstract}
In this paper, we discuss a voting model by considering three different kinds of networks: 
a random graph, 
 the Barab\'{a}si-Albert(BA) model, and a fitness model.
A voting model represents the way in which public perceptions are conveyed to voters.
Our voting model  is constructed by using two types of voters--herders and independents--and  two candidates.
Independents conduct voting  based on  their fundamental values;
on the other hand, 
herders base their voting on the number of  previous votes.
Hence, herders vote for the majority candidates and obtain information relating to previous votes from their networks.
We discuss the difference between the phases on which the networks depend.
Two kinds of phase transitions, an information cascade transition and a super-normal transition, were identified.
The first of these is a transition between a state in which most voters make the correct choices and a state in which most of them are wrong.
The second is a transition of convergence speed. 
The information cascade  transition  prevails when herder effects are stronger than the super-normal transition.
In the BA and fitness models, the critical point of the information cascade transition is the same as that of the random network model. However, the critical point of the  super-normal transition disappears when these two models are used.

In conclusion, the influence of networks is shown to only affect the convergence speed and not the information cascade transition. We are therefore able to conclude that the influence of hubs on voters' perceptions is limited.


\end{abstract}



\newpage
\section{Introduction}

Collective herding behavior has become a research topic in many research fields.
This kind of behavior is of interest in multi-disciplinary areas, such as sociology \cite{tarde}, social psychology \cite{mil},  ethnology \cite{fish},\cite{frank},  and economics. 
We consider statistical physics to be an effective tool for analyzing macro phenomena such as this. 
In fact,  the study of topics of this nature has led to the development of an associated  research field known as sociophysics \cite{galam}.  For example, in statistical  physics, anomalous fluctuations in financial markets \cite{Cont},\cite{Egu} and  opinion dynamics \cite{Stau},\cite{Curty},\cite{nuno},\cite{A}  have been discussed.
Recently, the effects of topologies on these dynamics has attracted considerable attention \cite{Cas},\cite{Mor}.

Human beings estimate public perception by observing the actions of other individuals, following which they exercise a choice similar to that of others.
This phenomenon is also referred to as social learning or imitation. 
Because it is usually sensible to do what other people are doing, collective herding behavior is assumed to be the result of a rational choice according to public perception. 
In ordinary situations this is the correct strategy.
However, this approach may lead to arbitrary or even erroneous decisions as a macro phenomenon, and 
is known as an  information cascade \cite{Bikhchandani}.

How do people obtain public perception?
In our previous paper we discussed the case in which people obtain information from previous $r$ voters using mean-field approximations \cite{Hisakado3}.
The model   is based on a one-dimensional (1D) extended lattice.
In the real world people obtain information from their friends and influencers.
A well-known example is the bunk run on the Toyokawa Credit Union in 1973. The incident was caused by a false rumor,
the process of which was subsequently analyzed by \cite{to1} and  \cite{to2}.
In fact, a rumor process such as this depends on individual relations.
The influencers became the hubs and affected many of those involved. 
Hence, in this work we consider a voting model in terms of different kinds of graphs, namely random graphs, the Barab\'{a}si-Albert(BA) model \cite{BA}, and the fitness model　\cite{BA1},\cite{BA2}, and compare the results to determine the effect of networks.

In our previous paper, we introduced a sequential voting model \cite{Hisakado2}, which
is based on a process in which one  voter  votes for one  of two  candidates at each time step $t$.
In the model, public perception is represented by allowing the  $t$-th voter to see all previous votes, i.e., $(t-1)$ votes.
Thus, there are two types of voters--herders and independents--and two candidates.

Herders' behavior is  known as the influence response function.
Threshold rules have been derived for a variety of relevant theoretical scenarios as the influence response function.
Some empirical and experimental evidence has confirmed the assumptions that individuals  follow threshold rules when making decisions in the presence of social influence \cite{watts2}.
This rule posits that individuals will switch between two choices only when a sufficient number of other persons have adopted the choice.
We refer to individuals such as these as digital herders.
From  our experiments, we  observed that 
human beings exhibit behavior between that of digital and analog herders \cite{Mori3}.
Analog herders vote for each candidate with probabilities that are proportional to candidates' votes.
In this paper,  we restrict the discussion to digital  herders to simplify the problem. 
\footnote{  The case of analog herders becomes the case of digital herders who refer one previous voter.}

Here,  we  discuss  a  similar voting model based on two candidates.
We define two types of voters--independents and herders.
Independent voters base their voting on their fundamental values and rationale.
They collect information independently.
On the other hand, 
herders exercise their voting based on the number of previous votes  candidates have received, which is visible to them  
in the form of public perceptions.
In this study, we consider the case wherein  a voter can see $r$ previous votes,  which depends on several graphs.

In the upper limit of $t$, the  independents cause the distribution of votes to converge to a  Dirac measure against herders. 
This model  contains two kinds of phase transitions.
The first is a transition of super and normal diffusions, and contains three phases--two super diffusion phases and a normal diffusion phase.
In super diffusion phases, the voting rate  converges  to  a Dirac measure  slower   than   in a  binomial distribution.
In a normal phase,  the voting rate converges as in a binomial distribution.
The other kind of phase transition  is referred to as an information cascade transition. 
As the fraction of herders increases, the model features a phase transition beyond which a state in which most voters make the correct choice coexists with one in which most of them are wrong.
This would cause the distribution of votes to change from one peak to two peaks. 
The purpose of our research is to clarify the way in which the network  of references  affects  these two phase transitions.




The remainder  of this paper is organized  as follows.
In section 2, we introduce our  voting model and 
  mathematically 
 define the two types of  voters--independents and herders. 
In section 3,
we derive a stochastic differential equation and discuss the voting model on the random graph.
In section 4, we discuss 
the voting model in terms of the BA model.
In section 5, we discuss the fitness model. 
This model uses hubs which affect the voters to a greater extent than the BA model. 
In section 6, we verify  these transitions through numerical simulations.
Finally, the conclusions are presented in section 7.

\section{Model}


We model the voting behavior of two candidates, $C_0$ and $C_1$,
at  time $t$, and $C_0$ and $C_1$ have  $c_0(t)$ and $c_1(t)$ votes, respectively.
In  each time step, one  voter  votes   for one  candidate, which means that the voting is sequential.
Hence,  at time $t$,  the $t$-th voter  votes,  after which the  total number of votes  is  $t$.
Voters are allowed to see  
$r$  previous votes for each candidate; thus, they are aware of   public perception.
Here $r$  is a constant number.
The selections that were made when $r$ previous votes were cast depend              
 on the networks to which voters have access.


In our model we assume an infinite number of voters of each of the two types--independents and herders.
The independents vote for candidates $C_0$ and $C_1$
with  probabilities $1-q$ and $q$, respectively.
Their votes are independent  of  others' votes, i.e., 
their votes are based  on  their fundamental values.
Here we assume $q\geq 1/2$.

On the other hand, the  herders' votes  are    based on the number of previous $r$ votes.
Here the voter  does not necessarily refer to the latest $r$ votes. 
We consider previous $r$ votes to mean those that were selected  by the voters' network.
Therefore, at time $t$, $r$ previous votes are 
the number of votes for $C_0$ and $C_1$, which is represented by $c_0^r
(t)$ and $c^r_1(t)$, respectively.
Hence, $c_0^r(t)+c_1^r(t)=r$ holds.
If $r > t$, voters can see $t$ previous votes for each candidate.
 In the limit $r \rightarrow \infty$, voters can see  all previous votes.
We define the number of  all previous votes for $C_0$ and $C_1$ as
 $c_0^\infty (t)\equiv c_0(t)$ and $c_1^\infty(t)\equiv c_1(t)$.
In the real world the number of references $r$ depends on the number of voters, but
here we specify $r$ to be constant.

In this paper, a herder is considered to be a digital herder \cite{Hisakado3},\cite{Hisakado2}.
Here we define $c(t)^r_1/r=1-c(t)^r_0/r=Z_r (t)$.
A herder's behavior is defined by the function $f(z)=\theta(Z_r-1/2)$, where
$\theta(Z)$ is a Heaviside function.



The independents and herders appear randomly and vote.
We set the ratio of independents to 
herders as $(1-p)/p$.
In this study, we mainly focus on  the upper  limit of $t$, which
refers to the voting of an infinite number of voters.

\begin{figure}[h]
\includegraphics[width=110mm]{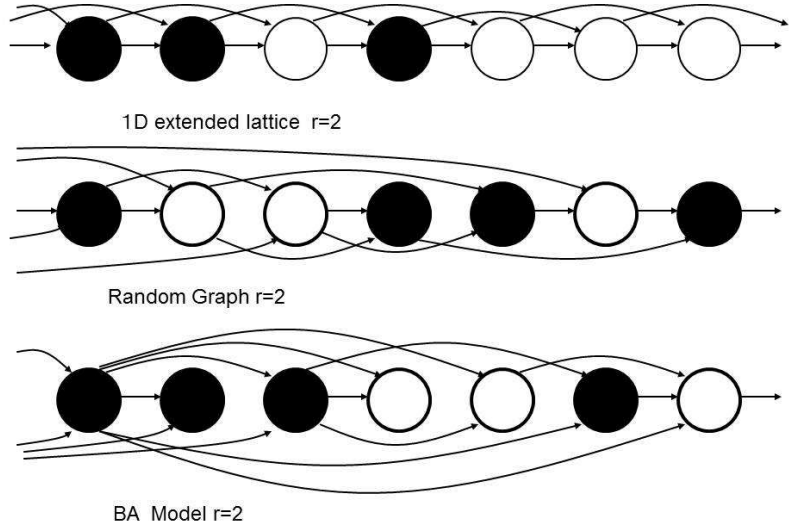}
\caption{Representation of graphs.
Shown is an extended 1D lattice, random graph, and BA  model when $r=2$.
The white (black) dot is a voter who voted for the candidate $C_0 (C_1)$. 
Two arrows pointing toward a dot represent a voter who refers to two voters when $r=2$.
In the case of an extended 1D lattice, a voter refers to the latest two voters.
In the case of a random graph, a voter  refers to two previous voters who are selected randomly.
In the case of the BA model, a voter refers to two previous voters who are selected via connectivity.
Hence, there are  voters who play the role of a hub in the BA model.
In the above figure, the first voter in the network graph is a hub who influences many other voters.}
\label{graph}
\end{figure}


We consider the voter to be able to see $r$ previous votes.
The problem becomes one of determining how the voter selects $r$ previous votes.
The influence of the reference voters is represented  as  a voting model on  networks.
How the network affects  the voting model is our problem. 
In this paper, we analyze cases by comparing models based on a  random graph, Barab\'{a}si-Albert(BA) model case, and  fitness model.
In Figure \ref{graph} we illustrate each one of these three cases for $r=2$.
We have previously discussed 1D extended lattice cases and showed that the analog and digital herder cases resemble the Kirman's ant colony and kinetic Ising models, respectively \cite{Hisakado4}.
In the figure, a white (black) dot is a voter who voted for candidate $C_0 (C_1)$. 
Two arrows point toward a dot, which means a voter refers to two other voters when $r=2$.
In the case of the 1D extended lattice, a voter refers to the latest two voters.
In the case of a random graph, a voter  refers two previous voters who are selected randomly.
In the case of the BA model,  a voter refers to two previous voters who are selected through the voter's connectivity network, which is introduced in the BA model \cite{BA}.
The BA model has the characteristics of a scale-free network with hubs.
The power index  of the BA model   is  three, whereas that of the fitness model  depends on the distribution of fitness.
In the case of a uniform distribution, it is 2.25, which  is below three.
Hence, there are  voters who play the role of a hub to connect other voters.
In the above figure, the first voter in the BA model corresponds to  a  hub.

\section{Random Graph}

We are interested  in the limit $t \rightarrow \infty$.  
At time $t$, the $t$-th voter selects $r$ voters who have already voted.
The $t$-th voter is able to see the total number of votes of selected different $r$ voters to obtain the information.
In this section we consider the case in which the voter selects 
$r$ votes randomly. 
Hence, this model is a voting model based on the random graph.

 
Herders vote for a  majority
candidate as follows.
If $c_0^{r}(t)>c_1^{r}(t)$, the majority of herders vote for the candidate $C_0$.
If $c_0^{r}(t)<c_1^{r}(t)$, the majority of herders vote for the candidate $C_1$.
If  $c_0^{r}(t)=c_1^{r}(t)$, herders vote for $C_0$ and $C_1$
with the same  probability, i.e.,$1/2$.
Here at time $t$, the selected  information of $r$ previous votes are 
the number of votes for $C_0$ and $C_1$, namely $c_0^r
(t)$ and $c^r_1(t)$, respectively.
The   herders in this section are known  as digital  herders.

We define $P_1^{r}(t)$ as the probability of the $t$-th voter voting for  $C_1$.
\begin{equation}
P_{1}^{r}(t)=
\begin{cases}
p+(1-p)q,\hspace{2cm}\,\,\,\,c_1^r(t)> r/2  \\
p/2+(1-p)q,\hspace{1.6cm}\,\,\,\, c_1^r(t)=r/2 \\
(1-p)q,\hspace{2.75cm} \,\,\,\,c_1^r(t)<r/2   \\
\end{cases}
\label {ge}
\end{equation}

In the scaling limit $t=c_0(t)+c_1(t)=c_0^{\infty}+c_1^{\infty}\rightarrow \infty$,
 we define
\begin{equation}
\frac{c_1(t)}{t}=Z(t)
\Longrightarrow Z_{\infty}.
\label{Z}
\end{equation}
$Z(t)$ is the ratio of voters who vote for $C_1$ at $t$.

Here we define $\pi$ as the majority  probability of binomial distributions of $Z$.
In other words, the probability of $c_1^r(t)>1/2$.
When  $r$ is odd,
\begin{equation}
\pi(Z)=\sum_{g=\frac{r+1}{2}}^{r}
\left(
\begin{array}{cc}
r\\
g  
\end{array}
\right)
Z^g(1-Z)^{r-g}\equiv\Omega_{r}(Z).
\label{2}
\end{equation}
When  $r$ is even, from the definition of the 
behavior of the herder,
\begin{eqnarray}
\pi(Z)&=&\sum_{g=\frac{r}{2}+1}^{r}
\left(
\begin{array}{cc}
r\\
g  
\end{array}
\right)
Z^g(1-Z)^{r-g}
+
\frac{1}{2}
\left(
\begin{array}{cc}
r\\
r/2  
\end{array}
\right)
Z^g(1-Z)^{r/2}
\nonumber \\
&=&
\sum_{g=\frac{r}{2}}^{r-1}
\left(
\begin{array}{cc}
r-1\\
g  
\end{array}
\right)
Z^g(1-Z)^{r-1-g}=\Omega_{r-1}(Z).
\label{3}
\end{eqnarray}
The majority probability $\pi$ in the even  case  becomes
the odd case $r-1$.
Hereafter we consider only the  odd case $r= 2n+1$, where $n=0,1,2, \dots$.

The value of $\pi$ can be calculated as follows,
\begin{equation}
\pi(Z)=\frac{(2n+1)!}{(n!)^2}\int_0^{Z}x^n(1-x)^ndx=\frac{1}{B(n+1,n+1)}\int_0^{Z}x^n(1-x)^ndx.
\label{pi}
\end{equation}
Equation (\ref{pi}) can be applied when the referred voters are selected to overlap.
In fact, in reality the referred voters are not selected to overlap.
However, in this paper  we use this approximation  to study a large $t$ limit.

We can rewrite  (\ref{ge}) for the random graph as
\begin{eqnarray}
c_1(t)&=&k \rightarrow k+1:
 P_{k,t}=p\pi(k/t)+(1-p)q,
\nonumber \\
c_1(t)&=&k   \rightarrow k:
 Q_{k,t}=1-P_{k,t},
\label{pd}
\end{eqnarray}

We define  a new variable $\Delta_t$ such that
\begin{equation}
\Delta_t=2c_1(t)-t=c_1(t)-c_0(t).
\label{d}
\end{equation}
We change the notation from $k$ to $\Delta_t$ for convenience.
Then, we have $|\Delta_t|=|2k-t|<t$.
Thus, $\Delta_t$ holds  within   $\{-t,t\}$. 
Given $\Delta_t=u$, we obtain a random walk model:
\begin{eqnarray}
\Delta_t&=&u \rightarrow u+1  :P_{u,t}=\pi(\frac{1}{2}+\frac{u}{2t})p+(1-p)q,
\nonumber \\
\Delta_t&=&u \rightarrow u-1  :Q_{u,t}=1-P_{u,t}.
\nonumber
\end{eqnarray}
We now  consider the continuous limit $\epsilon \rightarrow 0$,
\begin{equation}
X_\tau=\epsilon\Delta_{[\tau/\epsilon]},
\end{equation}
where $\tau=t\epsilon$.
Approaching the continuous limit, we can obtain the stochastic differential  equation (see Appendix A):
\begin{equation}
\textrm{d}X_\tau=[(1-p)(2q-1)-p+2p\frac{(2n+1)!}{(n!)^2}\int_0^{\frac{1}{2}+\frac{X_\tau}{2\tau}}x^n(1-x)^ndx ]\textrm{d}\tau+\sqrt{\epsilon}.
\label{ito0}
\end{equation}

In the case $r=1$, the equation becomes
\begin{equation}
\textrm{d}X_\tau=[(1-p)(2q-1)+p\frac{X_\tau}{\tau}]\textrm{d}\tau+\sqrt{\epsilon}.
\label{ito00}
\end{equation}
The voters vote for each candidate with probabilities that are proportional to the candidates' votes.
We refer to these herders as a kind of  analog herders \cite{Hisakado2}.

We are interested in the behavior at the   limit $\tau\rightarrow \infty$.
The relation between $X_{\infty}$ and the voting ratio to $C_1$ is $2Z-1=X_{\infty}/\tau$.
We consider the solution $X_\infty\sim\tau^{\alpha}$, where $\alpha\leq1$,  
since  the  maximum speed is $\tau$ when $q=1$.
The slow   solution is  $X_\infty\sim\tau^{\alpha}$, where $\alpha<1$ is hidden by
the fast    solution $\alpha=1$ in the upper  limit of $\tau$.
Hence, we  can assume a stationary solution as
\begin{equation}
X_\infty=\bar{v}\tau+(1-p)(2q-1)\tau,
\label{h}
\end{equation}
where $\bar{v}$ is constant.
Substituting  (\ref{h})  into (\ref{ito0}), we can obtain
\begin{equation}
\bar{v}=-p+\frac{2p\cdot(2n+1)!}{(n!)^2}\int_0^{\frac{1}{2}+\frac{(1-p)(2q-1)}{2}+
\frac{\bar{v}}{2}} x^n(1-x)^ndx.
\label{i}
\end{equation}
This is the  self-consistent equation.

Equation (\ref{i}) admits one solution  below the critical point $p\leq p_c$ 
 and  three  solutions for $p>p_c$.
When  $p\leq p_c$, we refer to the phase as the  one-peak phase.
When  $p>p_c$, the  upper  and lower solutions 
are  stable;
on the other hand, 
the intermediate solution is   unstable.
Then, the  two  stable solutions correspond to  good and bad equilibria, respectively,  
and  the distribution becomes the sum of the two Dirac measures.
We refer to this phase as a two-peaks phase.

If $r=2n+1\geq3$,  a phase transition occurs  in the range $0\leq p\leq1$.
If the voter obtains information from three of the above voters,
as the number  of herders increases, the model features   a phase transition beyond which  the state in which most voters make the correct  choice coexists with one in which most of them are wrong.
If the voter only obtains information from one or two other voters, there is no phase transition.
We refer to this transition as an information cascade transition \cite{Hisakado3}.


Next, we consider the phase transition resulting from convergence.
This type of transition has been  studied for analog herders \cite{Hisakado2}.
We expand $X_\tau$ around the solution $\bar{v}\tau+(1-p)(2q-1)\tau$.
\begin{equation}
X_\tau=\bar{v}\tau+(1-p)(2q-1)\tau+W_\tau.
\label{w}
\end{equation}
Here, we set $X_\tau\gg W_\tau$,
that is  $\tau\gg 1$.
We rewrite (\ref{ito0}) using (\ref{w}) and obtain the following:
\begin{eqnarray}
\textrm{d}W_\tau
&=&p\frac{(2n+1)!}{(n!)^2\cdot 2^{2n}}\frac{W_\tau}{\tau}[1-\{\bar{v}+(1-p)(2q-1)\}^2]^n\textrm{d}\tau+\sqrt{\epsilon}.
\label{ito3}
\end{eqnarray}
We use  relation (\ref{h}) and  consider  the first term of the expansion.
If we set $L=p(2n+1)!/\{(n!)^2\cdot 2^{2n}\}[1-\{\bar{v}+(1-p)(2q-1)\}^2]^n$, then (\ref{ito}) is identical to   (\ref{ito3}).

The phase transition of convergence is obtained from the explanation in Appendix B.
The critical point $p_{vc}$ is 
the solution of 
\begin{equation}
p\frac{(2n+1)!}{(n!)^2\cdot 2^{2n}}[1-\{\bar{v}+(1-p)(2q-1)\}^2]^n=\frac{1}{2},
\label{vc1} 
\end{equation}
and the self-consistent equation (\ref{i}).



Here we consider the symmetric case, $q=1/2$.
The self-consistent equation (\ref{i}) becomes
\begin{equation}
\bar{v}=-p+\frac{2p\cdot(2n+1)!}{(n!)^2}\int_0^{\frac{1}{2}+
\frac{\bar{v}}{2}} x^n(1-x)^ndx.
\label{ising12}
\end{equation}
The right-hand side (RHS) of (\ref{ising12})  rises  at $\bar{v}=0$.
If $r=1,2$, there is only one solution $\bar{v}=0$  in all regions of  $p$.
In this case, $Z$ only has one peak, at $0.5$, 
which indicates the one-peak phase.
In the case $r=1,2$, we do not observe an information cascade transition \cite{Hisakado3}.
On the other hand, in the case $r\geq3$, there are  two stable solutions and  an unstable solution $\bar{v}=0$ above $p_c$.
The   votes ratio for $C_1$ attains 
a good  or   bad equilibrium.
This is the so-called spontaneous symmetry breaking.
In one sequence, $Z$ 
 is  taken as $\bar{v}/2+1/2$ in the case of  a good equilibrium,  or
as  $-\bar{v}/2+1/2$ in the case of   a bad equilibrium,
where $\bar{v}$ is the solution of (\ref{ising12}).
This indicates  the two-peaks phase, and 
the critical point  is $p_c=\frac{(n!)^2}{(2n+1)!}2^{2n}$, where the gradient of the RHS of (\ref{ising12}) at $\bar{v}=0$  is $1$.
In the case of $r=3 (n=1)$, $p_c=2/3$ and $r=5(n=2)$, $p_c=8/15$.
As $r$ increases, $p_c$ moves towards $0$.
In the large limit $r\rightarrow \infty$, $p_c$ becomes $0$.
This is  consistent with the case in which herders obtain information from all previous voters.\cite{Hisakado}

Next, we consider the normal-to-super transition of the symmetric case $q=1/2$ by considering the case $r=2n+1\geq3$.
In this case, we observe  an information cascade transition.
If $r\leq 2$, we do not observe an  information cascade transition and   we can only observe  a part of the phases, as described  below.

In  the one-peak phase $p\leq p_c$, the only solution is  $\bar{v}=0$.
$p_c$ is the critical point  of the  information cascade transition. 
The first critical point of convergence  is $p_{vc1}=\frac{(n!)^2}{(2n+1)!}2^{2n-1}=\frac{p_c}{2}$.
When  $p\leq p_c$, $p_{vc1}$ is the solution of (\ref{ising12}) and (\ref{vc1}).
If  $0<p<p_{vc1}$, the voting rate for $C_1$ becomes $1/2$, and  the distribution converges as  in a binomial distribution.
If $p_{c}>p\geq p_{vc1}$,
   candidate $C_1$  gathers $1/2$ of all the  votes, also in the scaled distributions.
However, the rate at which the voting rate converges is lower than  in  a binomial distribution.
We refer to these phases as  super   diffusion phases.
There are two phases, $p=p_{vc1}$ and $p_{c}>p>p_{vc1}$ and they differ  in terms of their convergence speed.
If $r\leq 2$,  we can observe these three phases. 

Above $p_c$, in  the two-peaks phase,  we can obtain two stable solutions that are not $\bar{v}=0$.
At $p_c$, $\bar{v}$ moves from $0$ to one of these two stable solutions.
In one voting sequence, the votes converge to one of these stable  solutions.
If  $p_{c}<p\leq p_{vc2}$, the voting rate for $C_1$ becomes $\bar{v}/2+1/2$ or $-\bar{v}/2+1/2$, and  the convergence occurs  at a lower rate than in a binomial distribution. 
Here, $\bar{v}$ is the solution of (\ref{ising12}).
We refer to  this phase as a  super diffusion phase.
 $p_{vc2}$ is the second critical point of convergence from  the super to the normal diffusion phase,  and it is the solution of the simultaneous equations 
(\ref{ising12}) and (\ref{vc1}) when  $p> p_c$.
In the case $r=3$ we can obtain  $p_{vc2}=5/6$ at $\bar{v}=\pm\sqrt{(3p-2)/p}$. 
If  $p>p_{vc2}$, the voting rate for $C_1$ becomes $\bar{v}/2+1/2$ or $-\bar{v}/2+1/2$.
However, the distribution converges as   in a binomial distribution. This  is a  normal diffusion phase.
A total of  six  phases can be observed.

\section{Barab\'{a}si-Albert  model}

\begin{figure}[h]
\includegraphics[width=110mm]{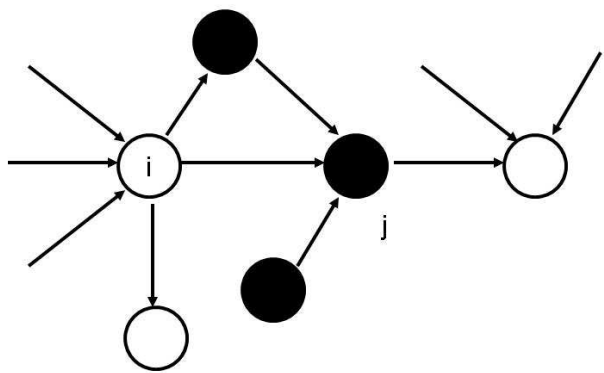}
\caption{The sample graph BA model with $r=3$. The voter $i$, who refers three voters and is referred by three voters, is
represented by a white circle, because they voted for candidate $C_0$. The connectivity of  $i$  is six.
 The voter $j$ refers three voters and is referred by a voter and is represented by a black circle, because of voting for candidate $C_1$. The connectivity  of $j$ is four.}
\label{net}
\end{figure}

In this section we consider the case in which the voter selects $r$ different  voters who 
will be selected based on popularity.
The popularity is proportional to the connectivity 
 of the voter $i$, such that
$l_i/\sum_j l_j $, where 
$l_i$ is the sum of the number of voters whom voter $i$ gave the information and from whom voter $i$ obtained  information, referenced against $r$ voters.
The total number of $l_i$  after the $t$-th voter has voted is $\sum_j l_j=2r(t-r+1)$, where $t\geq r$.
The sum $l_i$ corresponds to the connectivity in the BA model \cite{BA}. Hence, this model is the voting model based on the BA model.
The difference in the  connectivity is represented by different colors.
The color depends on whether the voter voted for candidate $C_1$ or $C_0$.
In Figure \ref{net}, voters who voted for $C_1$($C_0$) are represented by black(white) circles.
We define the   total connectivity number of voters who voted for candidate $C_1$($C_0$) at $t$ as $g_1(t)$($g_0(t)$).
Hence, $g_0(t)+g_1(t)=2r(t-r+1)$, where $t\geq r$.
In the scaling limit $g_0(t)+g_1(t)=2r(t-r+1)\rightarrow \infty$,
we define
\begin{equation}
\frac{g_1(t)}{2r(t-r+1)}=\hat{Z}(t)\Longrightarrow \hat{Z}_{\infty},
\end{equation}
 $\hat{Z}(t)$ is $g_1(t)/2r(t-r+1)$ at $t$.

We can write the evolution of connectivity as
\begin{eqnarray}
&  &g_1(t)=\hat{k} \rightarrow \hat{k}+i:
\nonumber \\
& (3r+1)/2\leq i\leq 2r& P_{\hat{k},t}(i)={}_{r}C_{2r-i}\hat{Z}^{i-r}(1-\hat{Z})^{2r-i}[(1-p)q+p],
\nonumber \\
& r+1\leq i\leq (3r-1)/2& P_{\hat{k},t}(i)={}_{r}C_{2r-i}\hat{Z}^{i-r}(1-\hat{Z})^{2r-i}(1-p)q,
\nonumber \\
&i=r&P_{\hat{k},t}(i)=(1-\hat{Z})^{r}(1-p)q+\hat{Z}^{r}(1-p)(1-q),
\nonumber \\
& (r+1)/2\leq i\leq r-1& P_{\hat{k},t}(i)={}_{r}C_{r-i}\hat{Z}^{i}(1-\hat{Z})^{r-i}(1-p)(1-q),
\nonumber \\
& 0\leq i\leq (r-1)/2& P_{\hat{k},t}(i)={}_{r}C_{r-i}\hat{Z}^{i}(1-\hat{Z})^{r-i}[(1-p)(1-q)+p].
\nonumber \\
\label{pd}
\end{eqnarray}

Here we consider the self-consistent equations for connectivity at a large $t$ limit.
\begin{equation}
2r\hat{Z}_{\infty}=\sum_{i=1}^{2r}P_{\hat{k},t}(i)\cdot i=r(1-p)q+rp\pi(\hat{Z}_{\infty})+r\hat{Z}_{\infty}.
\end{equation}
Hence, we can obtain
\begin{equation} 
\hat{Z}_{\infty}=(1-p)q+p\pi(\hat{Z}_{\infty}).
\label{sc1}
\end{equation}
On the other hand, the evolution  equation for the voting ratio $Z_{\infty}$ is
\begin{equation}
Z_{\infty}=(1-p)q+p\pi(\hat{Z}_{\infty}).
\label{sc2}
\end{equation}
Comparing (\ref{sc1}) and (\ref{sc2}), we  can obtain
$Z_{\infty}\sim \hat{Z}_{\infty}$. It means 
the behavior of the voting ratio, $Z_{\infty}$    is the same as that of the connectivity ratio, $ \hat{Z}_{\infty}$.\footnote{From (\ref{ito01}) the stochastic differential equation for the voting ratio becomes
\begin{eqnarray}
\textrm{d}X_\tau&=&[r(1-p)q-\frac{r}{2}+\frac{\hat{X}_{\tau}}{2(\tau-r+1)}
\nonumber \\
& &
+rp\frac{(2n+1)!}{(n!)^2}\int_0^{\frac{1}{2}
+\frac{\hat{X}_\tau}{2r(\tau-r+1)}}x^n(1-x)^ndx ]\textrm{d}\tau+\sqrt{\epsilon}.\nonumber
\end{eqnarray}(see Appendix A)}
Hereafter, we only analyze the behavior of the connectivity.

We define  a new variable $\hat{\Delta}_t$ such that
\begin{equation}
\hat{\Delta}_t=g_1(t)-r(t-r+1)=\frac{1}{2}\{g_1(t)-g_0(t)\}.
\label{d}
\end{equation}
We change the notation from $\hat{k}$ to $\hat{\Delta}_t$ for convenience.
Thus, $\hat{\Delta}_t$ holds  within   $\{-r(t-r+1),r(t-r+1)\}$. 
Given $\hat{\Delta}_t=\hat{u}$, we obtain a random walk model:
\begin{eqnarray}
&  &\hat{\Delta}=\hat{u} \rightarrow \hat{u}+i:
\nonumber \\
& (r+1)/2\leq i\leq r& P_{\hat{u},t}(i)={}_{r}C_{r-i}\hat{Z}^{i}(1-\hat{Z})^{r-i}[(1-p)q+p],
\nonumber \\
& 1\leq i\leq (r-1)/2& P_{\hat{u},t}(i)={}_{r}C_{r-i}\hat{Z}^{i}(1-\hat{Z})^{r-i}(1-p)q,
\nonumber \\
&i=0&P_{\hat{u},t}(i)=(1-\hat{Z})^{r}(1-p)q+\hat{Z}^{r}(1-p)(1-q),
\nonumber \\
& (-r+1)/2\leq i\leq -1& P_{\hat{u},t}(i)={}_{r}C_{-i}\hat{Z}^{r+i}(1-\hat{Z})^{-i}(1-p)(1-q),
\nonumber \\
& -r\leq i\leq -(r+1)/2& P_{\hat{u},t}(i)
={}_{r}C_{-i}\hat{Z}^{r+i}(1-\hat{Z})^{-i}[(1-p)(1-q)+p],
\nonumber \\
\label{pd2}
\end{eqnarray}
where $\hat{Z}=\hat{k}/(2r(t-r+1))=\hat{u}/(2r(t-r+1))+1/2$.


We now  consider the continuous limit $\epsilon \rightarrow 0$,
\begin{equation}
\hat{X}_\tau=\epsilon\hat{\Delta}_{[\tau/\epsilon]},
\end{equation}
where $\tau=t\epsilon$.
Approaching the continuous limit, we can obtain the stochastic  partial differential equation (see Appendix A):
\begin{eqnarray}
\textrm{d}\hat{X}_\tau&=&[r(1-p)q-\frac{r}{2}+\frac{\hat{X}_{\tau}}{2(\tau-r+1)}
\nonumber \\
& &
+rp\frac{(2n+1)!}{(n!)^2}\int_0^{\frac{1}{2}
+\frac{\hat{X}_\tau}{2r(\tau-r+1)}}x^n(1-x)^ndx ]\textrm{d}\tau+\sqrt{\epsilon}.
\label{ito0n}
\end{eqnarray}

For $r=1$, the equation becomes
\begin{equation}
\textrm{d}\hat{X}_\tau=[(1-p)(q-\frac{1}{2})+\frac{p+1}{2}\frac{\hat{X}_\tau}{\tau}]\textrm{d}\tau+\sqrt{\epsilon}.
\label{ito01}
\end{equation}
The voters  also vote for each candidate with probabilities that are proportional to   $\hat{X}_\tau$.\footnote{Here, we use the fact that the behavior of the voting ratio, $Z_{\tau}$    is the same as that of the connectivity ratio, $ \hat{Z}_{\tau}$. } However, equation (\ref{ito01}) is different from (\ref{ito00}).
The relation between  the voting ratio for $C_1$ and $\hat{X}_\infty$ is 
\begin{equation}
\frac{\hat{X}_\infty}{2r(\tau-r+1)}=\hat{Z}_{\infty}-
\frac{1}{2}.
\label{xz}
\end{equation}

We  can assume a stationary solution as
\begin{equation}
\hat{X}_\infty=r\bar{\hat{v}}\tau+r(1-p)(2q-1)\tau,
\label{hn}
\end{equation}
where $\bar{\hat{v}}$ is constant.
Since (\ref{xz}) and $0\leq\hat{Z}\leq1$, we can obtain
\begin{equation}
-1\leq\bar{\hat{v}}+(1-p)(2q-1)\leq1.
\label{hani}
\end{equation}
Substituting  (\ref{hn})  into (\ref{ito0n}), we can obtain
\begin{equation}
\bar{\hat{v}}=-p+\frac{2p\cdot(2n+1)!}{(n!)^2}\int_0^{\frac{1}{2}+\frac{(1-p)(2q-1)}{2}+
\frac{\bar{\hat{v}}}{2}} x^n(1-x)^ndx .
\label{i2}
\end{equation}
This is the  self-consistent equation and it agrees with (\ref{i}).
Then, in the BA model the information cascade transition behaves the same as in the random graph case. 
When  $p\leq p_c$, we refer to the phase as the  one-peak phase.
Equation (\ref{i2}) admits three  solutions for $p>p_c$.
When  $p>p_c$, the  upper  and lower solutions 
are  stable;
on the other hand, 
the intermediate solution is unstable.
The  two  stable solutions correspond to good and bad equilibria, respectively,  
and  the distribution becomes the sum of the two Dirac measures.
This is the two-peaks phase.
The phase transition point $p_c$ in the case of the BA model is the same as that in the random graph case.  
If $r=2n+1\geq3$,  a phase transition occurs  in the range $0\leq p\leq1$.
If the voter obtains information from either one or two voters, there is no phase transition.


Next, we consider the phase transition of convergence.
We expand $\hat{X}_\tau$ around the solution $r\bar{\hat{v}}\tau+r(1-p)(2q-1)\tau$.
\begin{equation}
X_\tau=r\bar{\hat{v}}\tau+r(1-p)(2q-1)\tau+r\hat{W}_\tau.
\label{w2}
\end{equation}
Here, we set $X_\tau\gg W_\tau$.
This indicates  $\tau\gg 1$.
We rewrite (\ref{ito01}) using (\ref{w2}) and obtain the following:
\begin{eqnarray}
\textrm{d}\hat{W}_\tau
&=&[1+p\frac{(2n+1)!}{(n!)^2\cdot 2^{2n}}(1-\{\bar{\hat{v}}+(1-p)(2q-1)\}^2)^n]\frac{W_\tau}{2\tau}\textrm{d}\tau+\sqrt{\epsilon}.
\label{ito4}
\end{eqnarray}
We use  relation (\ref{i2}) and  consider  the first term of the expansion.
If we set $L=1/2[1+p(2n+1)!/\{(n!)^2\cdot 2^{2n}\}(1-\{\bar{\hat{v}}+(1-p)(2q-1)\}^2)^n]$, (\ref{ito}) is identical to   (\ref{ito4}).

From Appendix B,
we can obtain   the phase transition of convergence.
The critical point $p_{vc}$ is 
the solution of 
\begin{equation}
[1+p\frac{(2n+1)!}{(n!)^2\cdot 2^{2n}}(1-\{\bar{\hat{v}}+(1-p)(2q-1)\}^2)^n]\frac{1}{2}
=\frac{1}{2},
\label{vc} 
\end{equation}
and (\ref{i2}).
The RHS of (\ref{vc}) is not less than $1/2$.
Using (\ref{hani})  we can obtain $p_{vc}=0$.
There is  no normal phase  in the case of the BA model.
The speed of convergence is always lower than normal in all regions of $p$.

\section{Fitness model}
In this section we consider the fitness model, which includes the BA model \cite{BA1}, \cite{BA2}.
We set the fitness of each voter equal to the weight of the connectivity. 
When the fitness has a constant distribution, the fitness model becomes the BA model.   
These  models lead to the appearance of stronger hubs.
The problem is whether the stronger hubs affect the phase transition points $p_c$ and $p_{vc}$.

The popularity is proportional to the weighted  connectivity 
 of voter $i$, such that
$\eta _il_i/\sum_j \eta_j l_j $, where
$l_i$ is the sum of the number of voters to whom voter $i$ gave the information and from whom   voter $i$ obtained  information, referenced against $r$ voters.
Further, $\eta_i$ is the fitness of voter $i$ and
 it is chosen from the distribution $u$.
The total number of weighted $l_i$  after the $t$-th voter has voted is $\sum_j l_j=2r(t-r+1)\bar{\eta}$, where $t\geq r$ and $\bar{\eta}$ is the average of the fitness.
When $u$ is a constant distribution, the fitness model becomes the BA model.
In the next section we discuss the cases for which the distribution  of  $\eta$ is constant, and uniform and exponential distributions for  numerical simulations.
We define the   total number  of weighted  connectivity  of the voters who voted for candidate $C_1$($C_0$) at $t$ as $g_1(t)$($g_0(t)$).
Hence, $g_0(t)+g_1(t)=2r\bar{\eta}(t-r+1)$, where $t\geq r$.
In the scaling limit $g_0(t)+g_1(t)=2\bar{\eta} r(t-r+1)\rightarrow \infty$,
we define
\begin{equation}
\frac{g_1(t)}{2r\bar{\eta}(t-r+1)}\Longrightarrow \hat{Z}_{\infty},
\end{equation}
where $\hat{Z}(t)$ is 
$g_1(t)/2r\bar{\eta} 
(t-r+1)$ at $t$.

Here we consider the self-consistent equations for connectivity in the large $t$ limit.
We set the  condition under which  voter $i$ votes and   $\hat{Z}_{\infty}$ does not change  as equilibrium.
Using (\ref{pd}) we obtain:
\begin{equation}
r\bar{\eta}\hat{Z}_{\infty}+r \eta_i \hat{Z}_{\infty}=r(1-p)q\eta_i+rp\pi(\hat{Z}_{\infty})\eta_i+r\bar{\eta}\hat{Z}_{\infty}.
\end{equation}
Hence, we can obtain
\begin{equation} 
\hat{Z}_{\infty}=(1-p)q+p\pi(\hat{Z}_{\infty}).
\label{sc11}
\end{equation}
which is the same as (\ref{sc1}).
This means that the transition point $p_c$ of the fitness model is that same as that of the BA model, for which
there is no super-normal transition.
From this result, we can estimate that there is no super-normal transition for the fitness models, details of which
are discussed in the next section. 



\section{Numerical Simulations}
In this section, we present a study of the voting model using several kinds  
of networks for which we use a numerical method. 
As fitness models,  we consider three types of  networks, which are 
constructed based on the mechanism of preferential attachment 
of nodes. The type of   the 
degree distribution is determined 
 by the distribution  $u$ of the fitness $\eta$  of each node.
We denote the networks  C,U, and E, 
as  the distribution $u$ is {\bf C}onstant, obeys {\bf U}niform 
distribution, and obeys {\bf E}xponential distribution.
C corresponds to the BA model.
In addition, we also study the voting models in a
{\bf R}andom graph and a 1D extended {\bf L}attice.
We name the networks L and R, respectively, and
adopt the model parameters $(r,q,p)$ as 
$r\in \{3,9\},q\in \{0.5,0.8\}$ and 
change the control parameter $p\in [0,1]$. 
We performed Monte Carlo simulations for time horizon $T$ as
$T=3.2\times 10^{4}$ and gathered $10^{5}$ samples. 
We calculated the average values of the macroscopic 
quantities with $S=10^{2}$ network samples.

The fitness model contains a phase transition similar to a Bose-Einstein  (BE) condensation \cite{BA2}.
The BE  condensation  represents  a "winner-takes-all" phenomenon for networks; in other words,
a few voters play the role of a large  hub.
The cases  C and U correspond to a scale-free phase and
the case E corresponds to a BE condensation phase.

\subsection{ Convergence of $Var(Z)$ }
\begin{figure}[htbp]
\begin{center}
\begin{tabular}{cc}
\includegraphics[width=7cm]{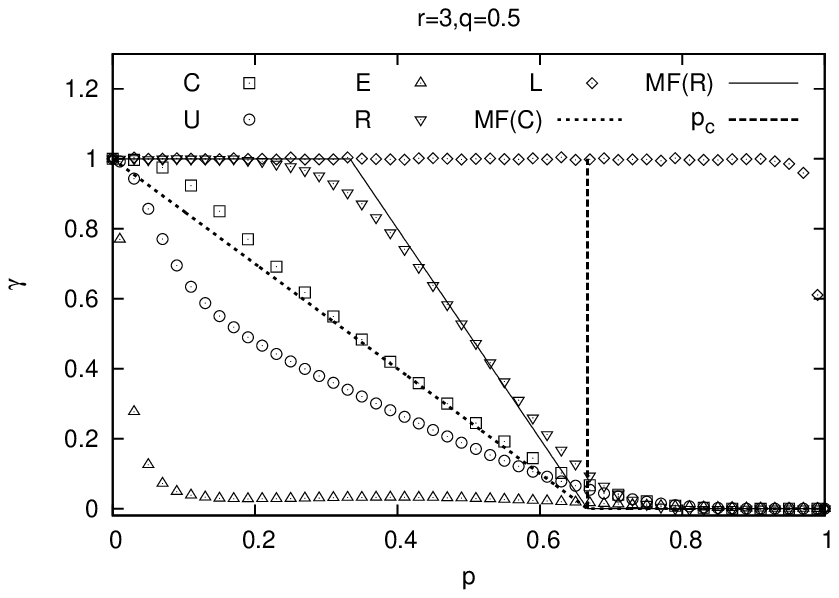} \
\includegraphics[width=7cm]{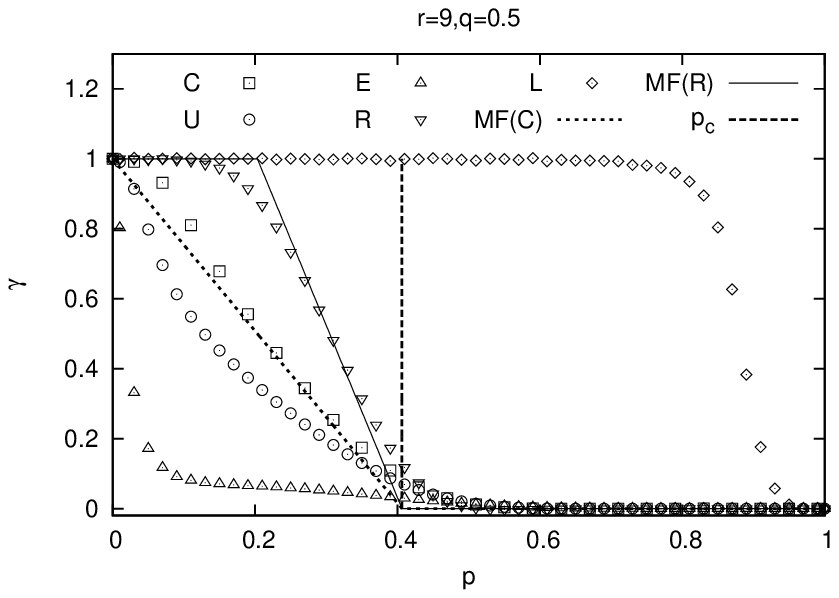} 
\vspace*{0.5cm} \\
\includegraphics[width=7cm]{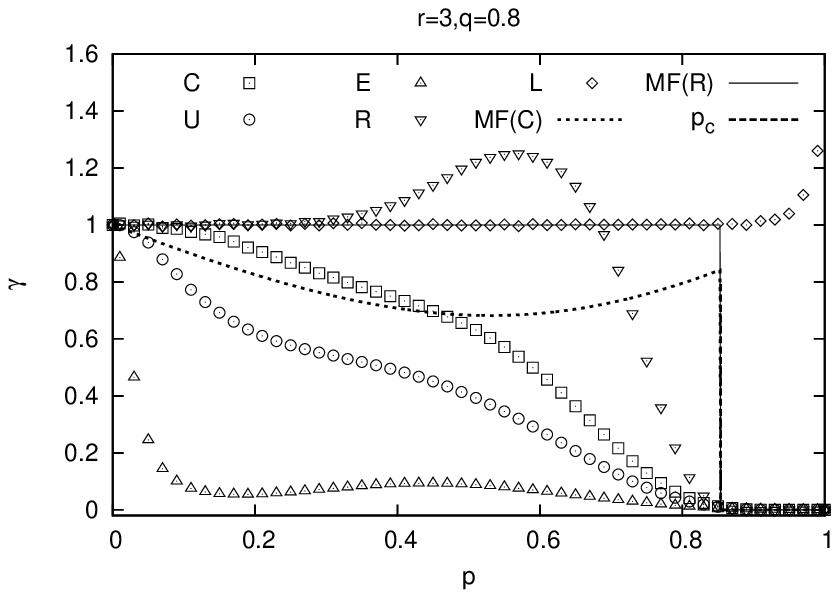} \
\includegraphics[width=7cm]{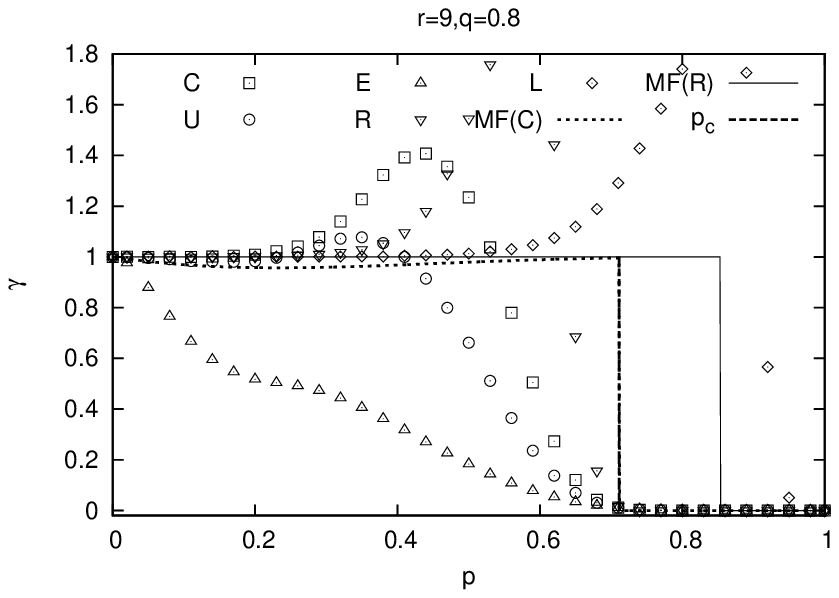} 
\end{tabular}
\end{center}
\caption{
Plot $\gamma$ vs. $p$. 
$q=0.5,r=3$ (Top Left),$q=0.5,r=9$ (Top Right),$q=0.8,r=3$ 
(Bottom Left) and $q=0.8,r=9$ (Bottom Right). 
$t=3.2\times 10^{4}$.
C($\Box$),U($\bigcirc$),E($\bigtriangleup$),R$(\bigtriangledown)$ and L($\diamond$).
The solid  and dotted  lines show the theoretical results of mean-field approximations for the R and C cases.
}
\label{syml}
\end{figure}

Figure \ref{syml} shows the simulations performed for the symmetric independent voter case, i.e., $q=1/2$.
Convergence  of distribution is observed in the  C, U, E,  L, and R  cases for the symmetric  case 
 $q=1/2$. 
MF is the theoretical line obtained in the previous sections.
The reference number  is  $r=3$ and  $r=9$.
Cases $q=0.5$ and $q=0.8$ represent independent voters.
The horizontal axis represents the ratio of herders $p$, and the vertical axis represents the speed of convergence $\gamma$.
We define $\gamma$ as $Var(Z)=\tau^{-\gamma}$, where $Var(Z)$ is the variance of $Z$. Here, $\gamma=1$ is the normal phase and $0<\gamma<1$ is the super diffusion phase. \footnote{Here, we estimate $\gamma$ from the slope of ${Var(Z(t))}$as $\gamma=\log [Var(Z(t-\Delta t))/Var(Z(t))]/\log[t/(t-\Delta t)]$.}

As discussed in the previous sections, the critical point at which the transition converges from normal diffusion to super diffusion is $p_{vc}\sim 0$ for the C, U, and E  cases.  
This means there is no  normal phase  for these cases.
These models  only have a super phase because of the hubs.
The case $L$ also does not have a phase transition from super to normal and only has a normal phase.
On the other hand, we can observe a phase transition from super to normal in the  R case.

At the critical point of the information cascade transition, the distribution splits into two delta functions and the exponent $\gamma$ becomes $0$.
For the C, U, E,  and R  cases, the critical point $p_c$ seems to be the same.
In the L case, there is no information cascade transition \cite{Hisakado4}.
In the next subsection, we discuss phase transitions  from the viewpoint of scaling.

.

\subsection{Estimator for $l$ and Var$(Z)$}
The phase transition of the voting models was detected by considering that the normalized 
correlation function $C(t)$ between the first voter's choice  
and $(t+1)$-th voter's choice plays a key role\cite{Mori:2014},\cite{Mori:2015}. 
The function is defined as the covariance divided by the variance of 
the first voter's choice.
The asymptotic behavior of $C(t)$ depends on the phase of the system
 and the expansion coefficient $l$.
In the one-peak phase, the leading term of $C(t)$ shows 
a power law dependence on $t$ as $t^{l-1}$.
In the two-peaks phase,
the limit value $c\equiv \lim_{t\to\infty}C(t)$ 
is positive and plays the role of the order parameter of 
the information cascade phase transition.
The sub-leading term depends on $t$ as $t^{l-1}$ and 
$l$ is the larger value $l_{max}$
of the two expansion coefficients $\{l_{+},l_{-}\}$ at the two peaks. 
We summarize the asymptotic behavior of $C(t)$ as
\begin{equation}
C(t) \simeq
\begin{cases}
a\cdot t^{l-1} ,\,\,\,\, c=0   \\
c+a\cdot t^{l-1} , \,\,\,\, c>0 . 
\label{eq:ct} 
\end{cases}
\end{equation}
Here, we write the coefficient of the term 
proportional to $t^{l-1}$ as $a$.
$c$ is equal to the limit 
value of $\lim_{t\to\infty}\mbox{Var}(Z(t))$ divided by 
the variance of the first voter's choice. 
On the phase boundary between the one-peak and two-peaks phase, 
the result for a generalized P\'{o}lya urn process suggests the following 
asymptotic behavior \cite{Mori:2015}.
\begin{equation}
C(t)\simeq c+a\cdot \log t^{-\alpha}+o(\log t^{-\alpha}) \label{eq:ct2}. 
\end{equation}
For the (a)symmetric case $q=\frac{1}{2}(\neq \frac{1}{2})$, 
$c=0 (>0)$ and $\alpha\simeq \frac{1}{2} (1)$.

To estimate $c$ and $l$, the integrated correlation 
time $\tau(t)$ and the second-moment correlation length $\xi(t)$ are useful. 
The values of the latter two parameters are defined through the $n$-th moment of $C(t)$ as,
\begin{equation}
\tau(t)\equiv M_{0}(t)\,\,\,\, , \,\,\,\, 
\xi(t) \equiv  
\sqrt{\frac{M_{n}(t)}{M_{0}(t)}}\,\,\,\, , \,\,\,\,
 M_{n}(t)\equiv \sum_{s=0}^{t-1}C(s)s^{n} \label{eq:def}.
\end{equation}
We estimate
the limit value of $\xi_{t}(t)\equiv \xi(t)/t$ 
and $\tau_{t}(t)\equiv \tau(t)/t$ using  
the asymptotic behavior of $C(t)$ in eq.(\ref{eq:ct}) as
\begin{equation}
\lim_{t\to \infty}\tau_{t}(t)=\lim_{t\to\infty}c+\frac{a}{l}t^{l-1}=c
\,\,\,\, , \,\,\,\,  
\lim_{t\to\infty}\xi_{t}(t)=
\begin{cases}
\sqrt{\frac{l}{l+2}},\,\,\,\, c=0   \\
\sqrt{\frac{1}{3}}, \,\,\,\, c>0 . 
\label{eq:limit} 
\end{cases}
\end{equation}
We can determine the one-peak phase or two-peaks phase 
by the limit value of $\tau_{t}(t)$.
In the one-peak phase, the limit value $\lim_{t\to \infty}\xi_{t}(t)$ can be used to
estimate $l$. 
On the phase boundary and in the two-peaks phase $l=1$.

\subsection{Order parameter $c$}

\begin{figure}[htbp]
\begin{center}
\begin{tabular}{cc}
\includegraphics[width=7cm]{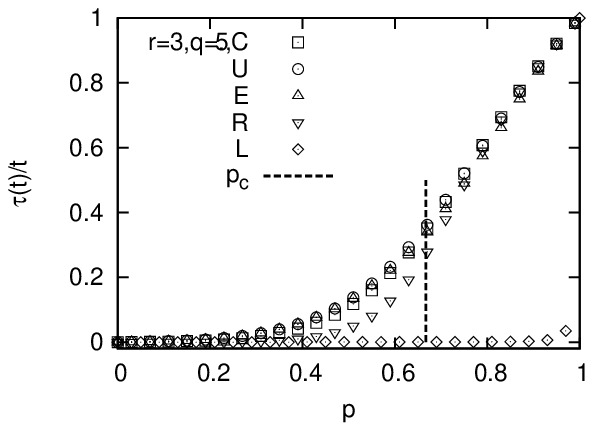} \
\includegraphics[width=7cm]{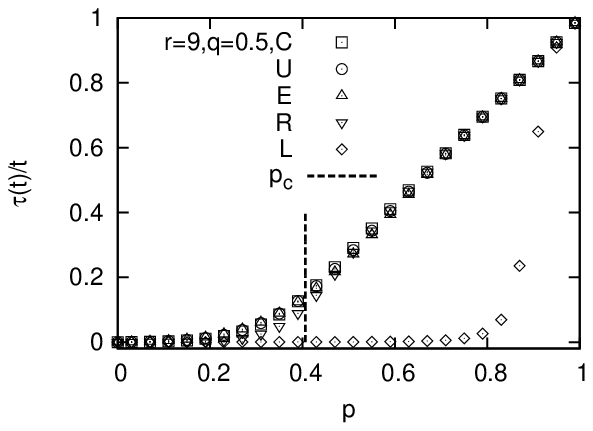} 
\vspace*{0.5cm} \\
\includegraphics[width=7cm]{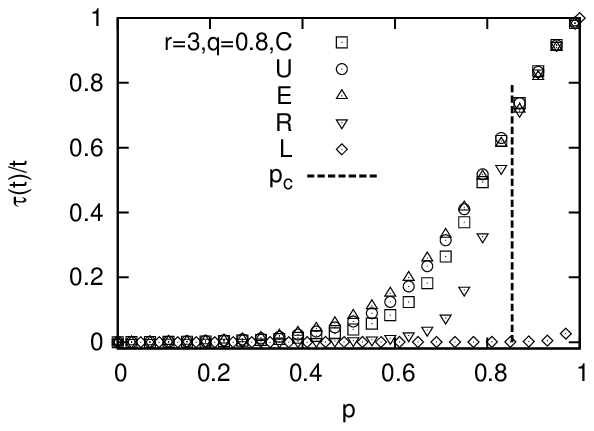} \
\includegraphics[width=7cm]{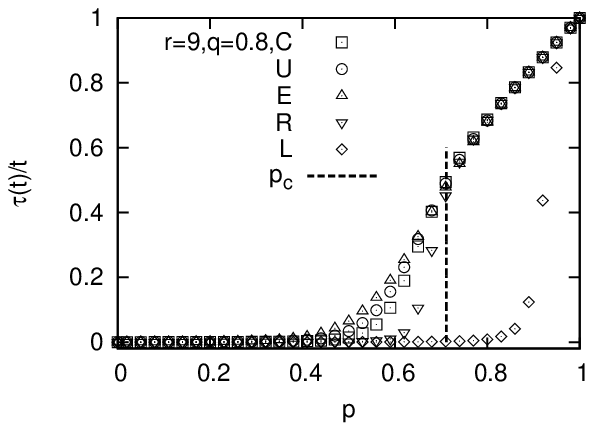} 
\end{tabular}
\end{center}
\caption{
Plot $\tau_{t}(t)$ vs. $p$. 
$q=0.5,r=3$ (Top Left),$q=0.5,r=9$ (Top Right),$q=0.8,r=3$ 
(Bottom Left) and $q=0.8,r=9$ (Bottom Right). 
$t=3.2\times 10^{4}$.
C($\Box$),U($\bigcirc$),E($\bigtriangleup$),R$(\bigtriangledown)$ and L($\diamond$).
The broken line shows the position of $p_{c}(q,r)$ for each case.
For the C, U, E, and R cases, $p_{c}(q,r)$ is the same.
In the L case, there is no phase transition. 
}
\label{fig:taut}
\end{figure}	

We estimate $\tau_{t}(t)$ for $t=T=3.2\times 10^{5}$ as a function of $p$ 
for C, U,  E, R, and L. The results 
for $r\in \{3,9\}$ and $q\in \{0.5,0.8\}$ are plotted in Figure \ref{fig:taut}.

In  the case L, $p_{c}=1$ and  $\tau_{t}(t)$ is almost zero for $p<1$ 
and becomes 1 at $p=1$.  
In other cases, $\tau_{t}(t)$ becomes an increasing function of $p$. 
The broken line shows the common 
position of $p_{c}$ in (\ref{i}) and (\ref{i2})  for C, U, E, and R.
If $p>p_{c}$, $\tau(t)/t$  is positive for $t=3.2\times 10^{4}$.
If $p$ is sufficiently small and $p<p_{c}$, $\tau(t)/t$ is almost zero.
In the region $p\simeq p_{c}$ and for $p<p_{c}$, 
 the limit value $\lim_{t\to\infty}\tau_{t}(t)$ should be zero as the system is
 in the one-peak phase. However, as can clearly be seen, 
the finite-size corrections are large and $\tau_{t}(t)$ assumes a large 
 positive value. To see the limit $\lim_{t\to \infty}\tau_{t}(t)$
 it is necessary to study the scale transformation $t\to \sigma t$ 
 and the limit  $\sigma\to \infty$.
The plots of $\tau_{t}(t)$ suggest 
that the value of $p_{c}$ is equal for C, U, E, and R. 

On the phase boundary $p=p_{c}$,
the situation is subtle and requires further careful study 
 to examine the limit value of $\tau_{t}(t)$. 
We think that the decrease of $\tau_{t}(t)$ with $t$ obeys a power 
law of $\log t$ as in eq.(\ref{eq:ct2}). A naïve extrapolation of the 
numerical results are not expected to work and it may be necessary to use finite-size scaling analysis 
to study the limit.

\subsection{Exponent $l$ in the one-peak phase}

We calculate $\xi(t)$ for $t=T$ and 
estimate $l$ by using the relation 
$\lim_{t\to\infty}\xi_{t}(t)=\sqrt{l/l+2}$. 
$l$ determines whether the system is in the one-peak phase 
or the two-peaks phase by the condition $l<1$ or $l=1$.
Furthermore, it also uses the condition $l >1/2$ or $l<1/2$ to determine whether the 
system is super-diffusive or normal diffusive.

As $\tau(t)\simeq ct+\frac{a}{l}t^{l}$ from eq.(\ref{eq:limit}), one
can derive the scaling relation for $\tau(t)$ under the 
scale transformation $t\to \sigma t$ as
\begin{equation}
\frac{\tau(\sigma t)}{\tau(t)}=\sigma^{l} \label{eq:scale}.
\end{equation}
In the two-peaks phase, we adopt $l=1$ in the relation.
Under the scale transformation, $\tau_{t}(t)$ does not change and 
the limit value $\lim_{t\to\infty}\tau_{t}(t)$ is positive if 
it is positive for sufficiently large $t$.
In the one-peak phase, $l<1$ and the limit value of 
$\tau_{t}(t)$ vanishes.

\begin{figure}[htbp]
\begin{center}
\includegraphics[width=9cm]{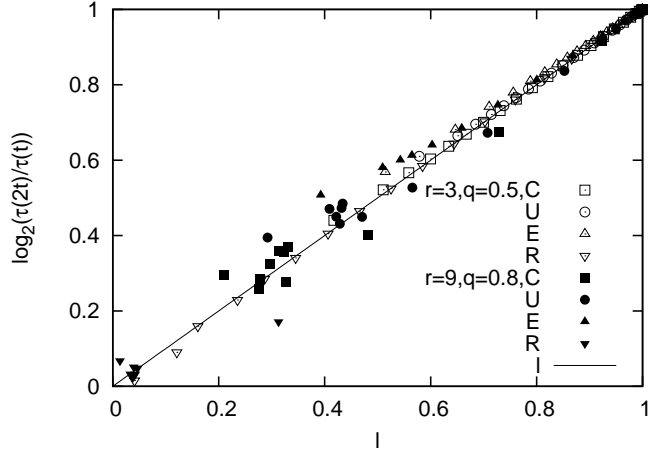} 
\end{center}
\caption{Plot of $\log_{2} \tau(2t)/\tau(t)$ vs. $l$.
$2t=T$,$(r,q)\in \{(3,0.5),(9,0.8)\}$ and C,U,E,and R networks.
}
\label{fig:rtau}
\end{figure}	
Figure \ref{fig:rtau} plots the logarithm of  
$\tau(2t)/\tau(t)$ with $2t=T$
as function of $l$. We use the data for 
$(r,q)\in \{(3,0.5),(9,0.8)\}$ and C,U,E,and R networks.
One sees that the data lie on the diagonal and the relation (\ref{eq:scale}) 
holds for the models in the C,U,E, and R networks. 
About L, $\tau(t)$ obeys another scaling relation \cite{Mori:2014}.

\begin{figure}[htbp]
\begin{center}
\begin{tabular}{cc}
\includegraphics[width=7cm]{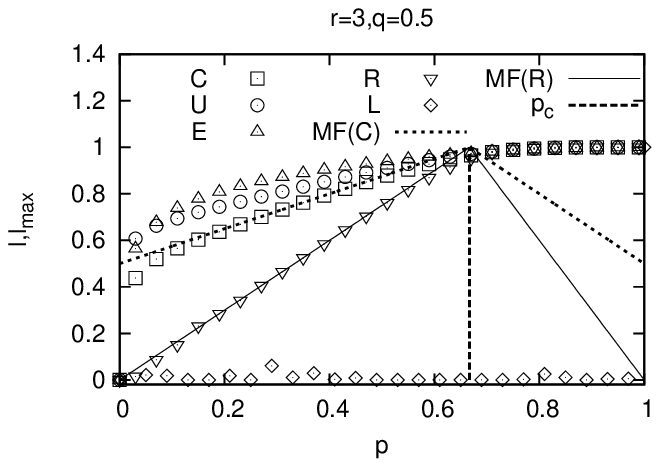} \
\includegraphics[width=7cm]{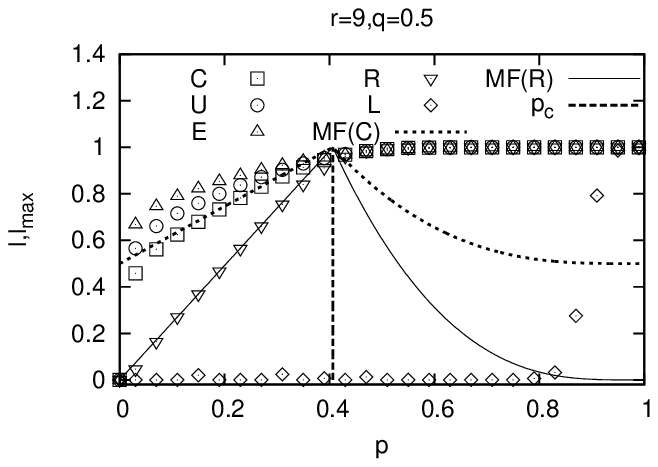}
\vspace*{0.3cm} \\
\includegraphics[width=7cm]{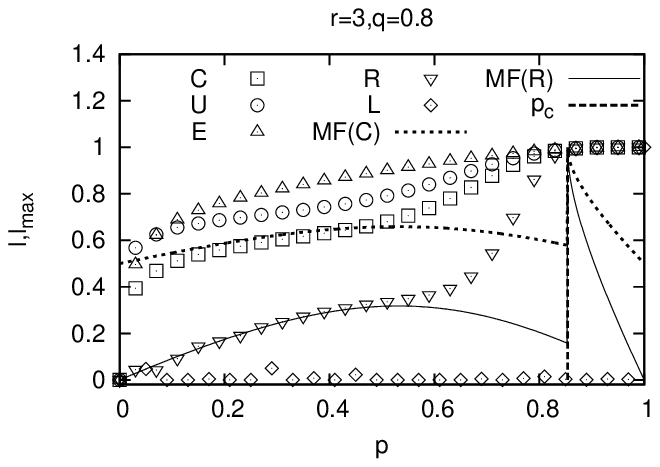} \ 
\includegraphics[width=7cm]{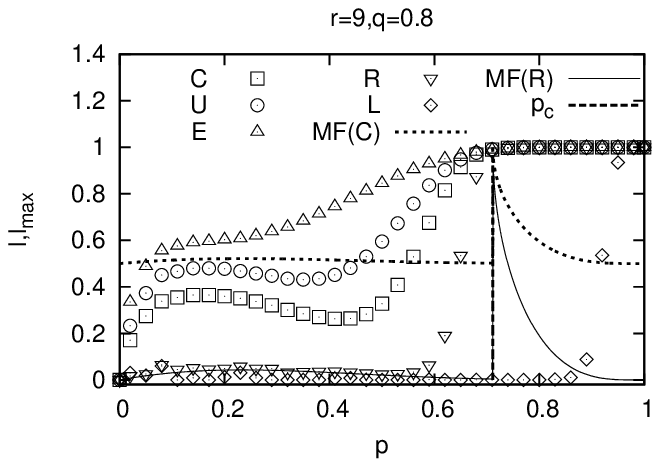} 
\end{tabular}
\end{center}
\caption{
Plot of $l$ estimated by $\xi_{t}(t)=\sqrt{l/(l+2)}$. 
The solid and dotted lines show the results of mean-field 
theory for $l_{max}$ in C and R networks, respectively.  
The broken line shows the position of $p_{c}$ for each case.
The parameters selections and the symbols are the same as those 
in Figure \ref{fig:taut}.
}
\label{fig:l}
\end{figure}	

Figure \ref{fig:l} plots $l$ as a function of $p$.
In L with $r=3$, $l=0$ for $p<p_{c}=1$ and 
the system is in the one-peak phase. 
$\tau_{t}$ are vanishingly small for $p<1$.
For $p\simeq 1$ with $r=9$, $l$ has a large positive value and 
touches 1 at $p<1$. This is an artefact of the finite-size correction.
If one estimates $l$ by using the large $T$ data, it should vanish for $p<1$.
 
The broken line shows the common position of $p_{c}$ for C, U,  E, and R networks.
We also plotted the theoretical results 
for $l$ in (\ref{i}),(\ref{i2})  by using thin solid and dotted lines for the C and R networks.
If $p>p_{c}$, $l$ is 1 and the systems are in the two-peaks phase.
If $p<p_{c}$, $l$ is less than 1 and it is in the one-peak phase.
The theoretical results for $l_{max}$ provide a good description of the behavior of $l$ 
in the one-peak phase.

For $q=0.5$ and C, $l$ is always larger than $0.5$ for $p>0$ and 
the system is in the super-diffusive phase.
In case R, $l$ changes continuously from 0 to 1 with $p$, which
 suggests the normal-diffusive phase transition.
For $q=0.8$, one see some discrepancy between the two 
estimates of $l$. It suggests that the time horizon 
$T=3.2\times 10^{4}$ is insufficiently large, preventing the system from reaching 
the scaling region. 
In case  C with $r=3$, $l>1/2$, which means that the system might be in the 
super-diffusive phase for $p>0$. In case C with $r=9$, we cannot deny the 
normal-diffusive phase with $l<0.5$. 
As the finite-size  correction is large for case R, 
we cannot deny the super-diffusive phase
 between the one-peak normal-diffusive phase and the two-peaks phase.
For U and E, the plots confirm that $p_{c}$ is common with C and R. 
As $l$  for U and E is always larger than that of C, the system might be in the 
one-peak super-diffusive phase for $p<p_{c}$. 
 
\section{Concluding Remarks}

In this paper we present a voting model  with collective herding behavior for the random graph, BA model, and fitness model   cases.
We investigated the phase differences between three different networks as models for the source of voter information.
This is based on the consideration that voters obtain their information from a network including hubs.
The BA and fitness models have networks that are similar  to real networks with hubs.
At the continuous limit, we could obtain  stochastic differential equations, which were subsequently used to 
analyze the difference between the models.

Our proposed model uses two kinds of  phase transitions, one of which is the
information cascade transition.
As the  number of herders increased, the model featured  a phase transition beyond which   a state in which most voters make the  correct choice coexisted with one in which most of them are wrong.
At this transition, the distribution 
of votes changed from the  one-peak phase to the  two-peaks phase.

The other phase transition  occurred at the convergence of the super and normal diffusions.
In  the  one-peak phase, a decrease in the number of herders caused the rate at which the variance  converged  to be slower than  in the binomial distribution.
This is the transition 
from normal diffusion to super diffusion.
This transition was also found in  the two-peaks phase,  in which sequential voting  converged to one of the two peaks.

In the case of the random graph model, all of these phases can be observed.
However, in the case of the BA and fitness models,
it is only possible to observe the super phase in the one-peak and two-peaks phases.
On the other hand,
in the case of the 1D extended lattice phase,
the only phase that exists is the one-peak  phase, which represents a normal convergence without 
any phase transition \cite{Hisakado4}.

The critical point $p_c$ is the same, regardless of whether the random graph, BA, or fitness models are used.
However, the difference  can be observed in the normal and super phases.
In the case of the BA model, there is only a super phase of which
the convergence speed is slower than for the normal phase.
In the case of the random  network model, the super phase and normal phase coexist.
The fitness model, which has stronger hubs than the BA model, has the same phase as the BA model. 
In conclusion, 
the influence of  hubs  can only be seen  in the convergence speed and  cannot  be seen in the phase transition between the one-peak and two-peaks phases.

In \cite{watts2}  the "influential hypothesis"  was discussed and negative opinions were expressed.
The hypothesis holds that influence, i.e., the existence of  hubs, is important for the formation of public opinion.
In our model the network does not affect  the critical point at which an information cascade transition occurs.
In this work hubs are only affected by the critical point  of super-normal
 transitions for a large $t$ limit.
The phase transition is the  transition of the speed of convergence.
Hence, although the existence of hubs   affects   the standard deviation of the votes, they are unable to change the  final outcome.
Therefore,  we can conclude that a hub has limited influence.

In this paper we discussed the case that herders’  threshold  is one half, i.e., that there is an equal probability of herders voting for either one of two candidates. 
 To confirm our conclusions,  we will make  a two-choice quiz experiment on networks which is similar to  \cite{Mori3}. 
 In the real world, the threshold  is not a half.  We discussed the bunk run case of Toyokawa Credit Union \cite{to1}, \cite{to2}.  In this case we are required to go to  a bank without immediate confirmation.  Because we are sensitive about rumors such as this, the threshold is reduced to  much below a half.  In next paper, we plan to study the case for which the threshold  is a variable and compare   to   observations \cite{nuno}, \cite{A}.    

Considering the intermediate  case between a 1D extended lattice and a random network,
in the previous paper,
 we showed that a 1D extended lattice is characterized by the absence of a phase transition and the presence of a one-peak phase \cite{Hisakado4}.
In contrast, a random network has both information cascade and super-normal transitions.
Investigating intermediate phases would be an interesting problem.
These  networks are nothing but small-world networks \cite{WS}.
Investigation of a voting model on a small-world network   is  a  future problem.

\appendix
\def\thesection{Appendix \Alph{section}}
\section{Derivation of stochastic differential equation} 

We use  $\delta X_\tau=X_{\tau+\epsilon}-X_\tau$ and $\zeta_\tau$, a standard i.i.d. Gaussian sequence with the objective to identify the drift $f_\tau$ and  the variance $g^2_\tau$  such that
\begin{equation}
\delta X_\tau=f_\tau(X_\tau)\epsilon+\sqrt{\epsilon}g_\tau(X_\tau)\zeta_{\tau+\epsilon}.
\end{equation}
Given $X_\tau=x$, using the transition probabilities of $\Delta_n$, we obtain
\begin{eqnarray}
\textrm{E}(\delta X_\tau)&=&\epsilon \textrm{E}(\Delta_{[\tau/\epsilon]+1}-\Delta_{[\tau/\epsilon]})=\epsilon(2p_{[\frac{l/\epsilon+\tau/\epsilon}{2}],\tau/\epsilon}-1)
\nonumber \\
&=&
\epsilon[(1-p)(2q-1)-p+2p\frac{(2n+1)!}{(n!)^2}\int_0^{\frac{1}{2}+\frac{X_\tau}{2\tau}}x^n(1-x)^ndx ].
\end{eqnarray}
Then, the drift term is $f_\tau(x)=(1-p)(2q-1)+p 
\tanh  (\lambda x/2\tau)$.
Moreover, 
\begin{equation}
\sigma^2(\delta X_\tau)=\epsilon^2
[
1^2p_{[\frac{l/\epsilon+\tau/\epsilon}{2}],\tau/\epsilon}
+(-1)^2(1-p_{[\frac{l/\epsilon+\tau/\epsilon}{2}],\tau/\epsilon})]
=\epsilon^2,
\end{equation}
  such  that 
$g_{\epsilon,\tau}(x)=\sqrt{\epsilon}.$
We can   obtain  $X_\tau$ such that it  obeys a diffusion equation with small additive noise:
\begin{equation}
\textrm{d}X_\tau=[(1-p)(2q-1)-p+2p\frac{(2n+1)!}{(n!)^2}\int_0^{\frac{1}{2}+\frac{X_\tau}{2\tau}}x^n(1-x)^ndx ]\textrm{d}\tau+\sqrt{\epsilon}.
\label{ito}
\end{equation}

\section{Behavior of solutions   of the stochastic differential equation}

We  consider the stochastic differential equation
\begin{equation}
\textrm{d}x_\tau=
(\frac{L x_\tau}{\tau})\textrm{d}\tau+\sqrt{\epsilon},
\label{ito}
\end{equation}
where $\tau\geq1$.

Let $\sigma^2_1$
be the variance of $x_1$.
If $x_1$ is Gaussian   $(x_1\sim\textrm{N}(x_1,\sigma^2_1))$ or deterministic $(x_1\sim\delta_{x1})$, the law of $x_\tau$ ensures that the  Gaussian is in  accordance with density
\begin{equation}
p_\tau(x)\sim
\frac{1}{\sqrt{2\pi}\sigma_\tau}\textrm{e}^{-(x-\mu_\tau)^2/2\sigma_\tau^2},
\end{equation}
where $\mu_\tau=\textrm{E}(x_\tau)$ is the expected value of $x_\tau$ and 
$\sigma^2_\tau\equiv \nu_\tau$ is its variance.
If $\Phi_\tau(\xi)=\log(\textrm{e}^{\textrm{i}\xi x_\tau})$
 is the logarithm of the characteristic function of the law of $x_\tau$, we have\begin{equation}
\partial_\tau
\Phi_\tau(\xi)
=\frac{L}{\tau}\xi\partial_\xi\Phi_\tau(\xi)
-\frac{\epsilon}{2}\xi^2,
\end{equation}
and
\begin{equation}
\Phi_\tau(\xi)=\textrm{i}\xi \mu_\tau-\frac{\xi^2}{2}\nu_\tau.
\end{equation}
Identifying
the real and imaginary parts of $\Phi_\tau(\xi)$, we 
obtain the dynamics of  $\mu_\tau$ as
\begin{equation}
\dot{\mu}_\tau=\frac{L}{\tau}\mu_\tau.
\end{equation}
The solution for $\mu_\tau$ is
\begin{equation}
\mu_\tau=x_1\tau^L.
\end{equation}
The dynamics of $\nu_\tau$ are  given by the Riccati equation
\begin{equation}
\dot{\nu}_\tau=\frac{2L}{\tau}\nu_\tau+\epsilon.
\label{gp}
\end{equation}
If $\nu\neq 1/2$, we get
\begin{equation}
\nu_\tau=
\nu_1\tau^{2L}+\frac{\epsilon}{1-2L}(\tau-\tau^{2L}).
\end{equation}
If $l=1/2$, we get
\begin{equation}
\nu_\tau=\nu_1\tau+\epsilon\tau\textrm{log}\tau.
\end{equation}
We can  summarize the temporal behavior of the variance as 
\begin{equation}
\nu_\tau\sim\frac{\epsilon}{1-2L}\tau\hspace{1cm}\textrm{if}\hspace{0.5cm}L<\frac{1}{2},
\end{equation}
\begin{equation}
\nu_\tau\sim(\nu_1+\frac{\epsilon}{2L-1})\tau^{2L}\hspace{1cm}\textrm{if}\hspace{0.5cm}L>\frac{1}{2},
\end{equation}
\begin{equation}
\nu_\tau\sim\epsilon\tau\textrm{log}(\tau)\hspace{1cm}\textrm{if}\hspace{0.5cm}L=\frac{1}{2}.
\end{equation}


This model has three phases. If $L>1/2$ or $L=1/2$,
    $x_\tau/\tau$ converges  slower than  in  a binomial distribution.
These phases  are the super diffusion phases.  
If  $0<p<1/2$, $x_\tau/\tau$   converges as    in a binomial distribution.
This is the normal phase \cite{Hisakado2}.

\section{Mean field approximation and Stochastic differential equation }

Here we discuss the relation between the stochastic differential equation and mean-field approximation.

\begin{figure}[h]
\includegraphics[width=140mm]{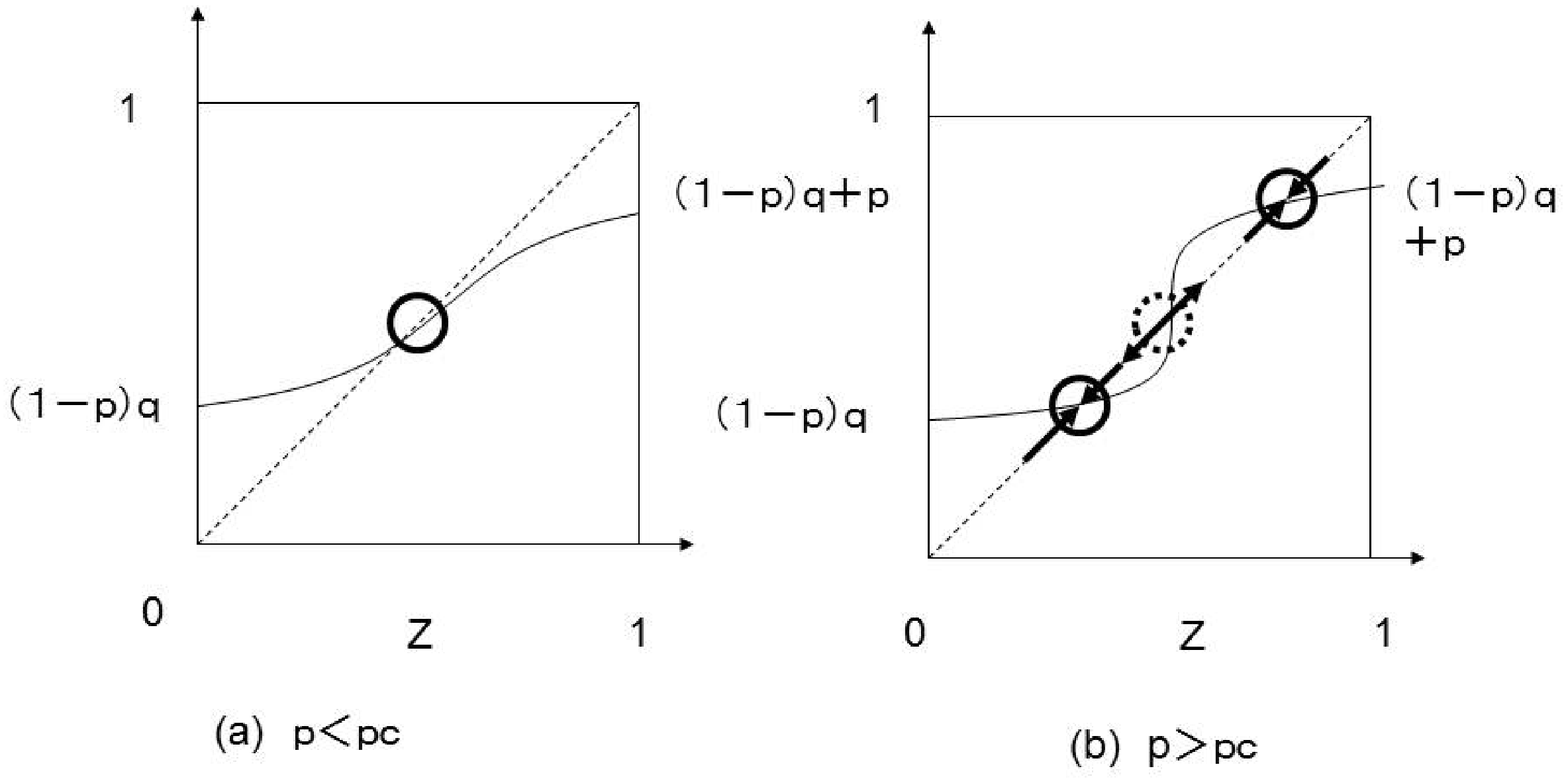}
\caption{Solutions of self-consistent equation (\ref{mf2}).  (a) $p\leq p_c$  and (b) $p>p_c$. Below the critical point $p_c$, we can obtain one solution (a). We refer to  this phase  as the  one-peak phase. In contrast, above the critical point, we obtain three solutions. Two of them are stable and one is unstable (b). We refer to  this phase as the  two-peaks phase.}
\label{Ising}
\end{figure}

At first we discuss the random graph case.
The relation between $X_{\infty}$ and the voting ratio to $C_1$ is $2Z_{\infty}-1=X_{\infty}/\tau$.
Hence, we can obtain from (\ref{h})
\begin{equation}
2Z_{\infty}-1=\bar{v}+(1-p)(2q-1).
\label{mf}
\end{equation}
Substituting (\ref{mf}) into (\ref{i})
 we can obtain
\begin{equation}
Z_{\infty}=q(1-p)+\frac{p\cdot(2n+1)!}{(n!)^2}\int_0^{Z_{\infty}} x^n(1-x)^ndx=q(1-p)+p\pi(Z_{\infty}). 
\label{mf2}
\end{equation}
The first term of (\ref{mf2}) is the contribution from independent voters and 
the second term is from herders as the sum of probabilities of every combination of majorities in the difference of previous $r$ voters.
In Fig\ref{Ising}, we show the relations between (\ref{mf2}) and information phase transition. 
 Below the critical point $p_c$, we can obtain one solution Fig\ref{Ising} (a). We refer to  this phase  as the  one-peak phase. In contrast, above the critical point, we obtain three solutions. Two of them are stable and one is unstable Fig\ref{Ising} (b). We refer to  this phase as the  two-peaks phase.

For the network case, the self-consistent equations for the network case
(\ref{i2}) and  for the random  graph case (\ref{i}) are the same.
Hence, we can obtain the mean-field approximation for the connectivity
as (\ref{sc1}).

\end{document}